\documentclass[
 aip,
 jmp,%
 amsmath,amssymb,
superscriptaddress,
reprint,
floatfix
]{revtex4-1}
\usepackage{comment}
\usepackage{graphicx}
\usepackage{dcolumn}
\usepackage{bm}
\usepackage{textcomp}
\usepackage{amsmath}
\usepackage{multirow}
\usepackage{hyperref}
\usepackage{xcolor}
\hypersetup{
    colorlinks=true,
    linkcolor=blue,
    citecolor=blue,
    urlcolor=cyan,
}
\begin{document}

\preprint{AIP/123-QED}

\title{Tunable Skyrmion-Antiskyrmion Dynamics in Co/Pt Nanocontacts for Spintronic Applications\\
}

\author{Hind Prakash}
\affiliation{Department of Physics, Indian Institute of Technology Roorkee, Roorkee 247667, India.}
\author{Himanshu Fulara}
\email{Author to whom correspondence should be addressed: himanshu.fulara@ph.iitr.ac.in}
\affiliation{Department of Physics, Indian Institute of Technology Roorkee, Roorkee 247667, India.}

\date{\today}

\begin{abstract}
Magnetic skyrmions are topologically protected quasiparticles and have drawn much attention because of their potential applications in next-generation spintronics devices. Their inherent topological stability, nanoscale size, and efficient manipulation via spin currents make them promising candidates for high-density data storage and novel computing paradigms. We micromagnetically investigate the nucleation dynamics of magnetic skyrmion pairs excited underneath two 30 nm nanocontacts with varying separations on top of an extended Co/Pt bilayer thin film. At close separation of 100 nm, the magnetization configurations strongly interact, giving rise to the formation of stable merged skyrmion states. As the separation increases beyond 200 nm, topologically distinct metastable configurations emerge, including the coexistence of tunable skyrmion-antiskyrmion pairs through Dzyaloshinskii-Moriya interaction (DMI) strengths and current pulse amplitudes. These metastable states eventually relax into two stable skyrmions that can be independently toggled ON and OFF using a weak in-plane magnetic field, enabling complex logic operations and more flexible circuit designs. Beyond the fundamental interest in skyrmion interaction dynamics, the independent control of skyrmion-antiskyrmion states holds promise for next-generation spintronic devices, with potential applications in memory, logic, and computing.
\end{abstract}


\maketitle

Magnetic skyrmions are nanoscale spin structures with topological stability~\cite{Roessler2006nt,Yu2010nt,marrows2021apl,lee2023apl}, often nucleated in materials exhibiting either interfacial or bulk Dzyaloshiskii-Moriya interaction (DMI). Since their first experimental observation in 2009~\cite{Muhlbauer2009sc}, spin-transfer (STT) and spin-orbit torque (SOT)-driven magnetic skyrmions~\cite{jonietz2010sc,iwasaki2013,jiang2015sc,woo2016ntm} have garnered significant attention due to their potential use in next-generation energy-efficient spintronics devices~\cite{marrows2021apl}. These applications range from high-density data storage~\cite{fert2013ntn,sampaio2013ntn,tomasello2014scirep} and logic operations~\cite{zhang2015sr,sisodia2022pra} to neuromorphic computing~\cite{grollier2020natel,song2020ne,marrows2024npjs} and quantum computing as skyrmion qubits~\cite{Psaroudaki2021prl}. The deformability of magnetic skyrmions makes them particularly promising for emerging technologies such as reservoir computing~\cite{Prychynenko2018pra,lee2023apl} and random number generation~\cite{wang2022ntc}. 

In racetrack memory devices~\cite{fert2013ntn,tomasello2014scirep} based on ferromagnetic skyrmions as information carriers, data is typically encoded through the presence (``1") or absence (``0") of a skyrmion, which requires precise control of the skyrmion's position. However, achieving this precision is often hindered by the skyrmion Hall effect~\cite{nagaosa2013ntn,jiang2017ntp}, which causes skyrmions to drift toward device edges, resulting in unwanted annihilation. To address this challenge, it is highly advantageous to investigate other localized metastable spin structures, such as skyrmionium or the coexistence of the skyrmions-antiskyrmions phase, as potential bits~\cite{goerzen2023npjqm}. Systems capable of hosting multiple coexisting metastable spin structures are not only of fundamental interest but also hold significant promise for advanced technological applications. For practical implementation, the controllable nucleation of skyrmions is essential, since nucleation mechanisms and their intermediate states play a critical role in determining skyrmion stability. Among the various methods proposed for skyrmion nucleation~\cite{sampaio2013ntn,woo2016ntm,durrenfel2017prb,buttner2017ntn, legrand2017nanolett,everschor2017newjphy,soumyanarayanan2017ntm,kern2022nanolett,je2018nanolett,lemesh2018advmater,marrows2021apl}, nanocontact (NC)-based nucleation using spatially nonuniform SOT has emerged as a promising energy-efficient approach. In a single nanocontact, SOT-induced skyrmion nucleation is often preceded by the skyrmionium~\cite{durrenfel2017prb,bogdanov1999jmmm,zhang2016prb,gobel2019scirep,bo2020jopdap,yang2023ntc,nakamura2024prb,qiu2024alp} state, which acts as an intermediate state.  However, in devices with multiple NCs, the metastable states can vary depending on the interactions among complex magnetic domains. Thus, a comprehensive understanding of these metastable configurations in multi-nanocontact devices is of growing interest for optimizing skyrmion-based memory and logic applications.

This study investigates the nucleation and interaction dynamics of skyrmion pairs in Co/Pt-based NCs, demonstrating the conditions that enable the generation and manipulation of topologically distinct metastable states and skyrmion pairs. Understanding these dynamics is essential for optimizing skyrmion-based devices, as our findings indicate that interactions between metastable states can lead to complex phenomena, such as the merging of magnetic domains into a single skyrmion, the emergence of skyrmion-antiskyrmion pairs, and the stabilization of skyrmion pairs.  By examining the impact of key parameters—including NC separation, DMI strength, and weak in-plane magnetic fields—this work provides a comprehensive understanding of the mechanisms driving skyrmion pair formation and their behavior in confined geometries.
 
\begin{figure*}[t]
\centering
\includegraphics[width=16cm]{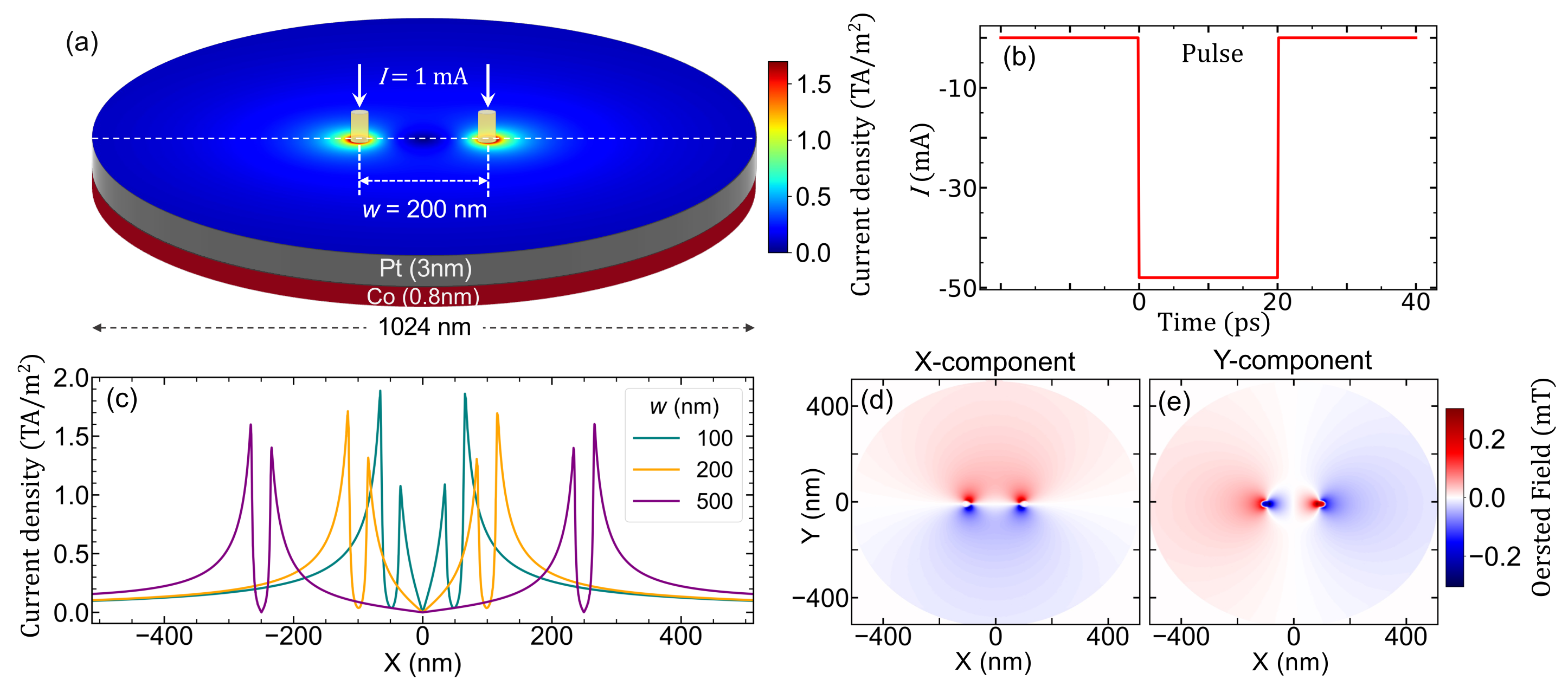}
\caption{Device schematic, current density, and Oersted field distribution. (a) Schematic representation of two NCs separated by 200 nm, with colors indicating the in-plane current density distribution in the Pt layer. (b) Application of a -48 mA square current pulse with a width of 20 ps. (c) Distribution of the current density in the plane along the diameter of the disc intersecting the centers of NCs (indicated by the white dashed line in (a)) for separations of 100 nm, 200 nm, and 300 nm. (d),(e) x- and y-components of the Oersted field distribution in the Co layer for NCs with a 200 nm separation under a 1 mA current.}
\label{fig:1} 
\end{figure*}

We modeled the device geometry using COMSOL Multiphysics~\cite{COMSOL}, consisting of two 30 nm diameter nanocontacts (NCs) on an extended Pt(3 nm)/Co(0.8 nm) bilayer, with separations varying from 100 nm to 500 nm. Figure \ref{fig:1}(a) shows the representative current density distribution at the midplane of the Pt layer for a 200 nm NC separation under an applied current of 1 mA, alongside the device schematic. The resistivity values used were 12 $\mu\Omega$ cm for Pt and 40 $\mu\Omega$ cm for Co~\cite{Li2000vacsci}. Due to its higher resistivity compared to Pt, the current in Co was minimal and thus neglected in the simulations. The current density was found to peak near the edges of the NCs, displaying an asymmetrical distribution around them. Figure \ref{fig:1}(b) illustrates the shape and amplitude of the -48 mA threshold current pulse used to nucleate skyrmions for the 200 nm NC separation. Figure \ref{fig:1}(c) shows the in-plane current density profile along a line passing through the centers of both NCs (indicated by the white dashed line), for an applied current of 1 mA. The current density reached its maximum at the outer peripheries of the NCs and increased as the NC separation decreased, while being minimal at the center of the disc. The spin current injected into the Co layer, denoted by $J_S$, is proportional to the in-plane current density $J_C$ in the Pt layer, expressed as $J_S = \Theta_{SH} J_C$, where $\Theta_{SH}$ is the spin Hall angle of Pt, taken as 0.08~\cite{ando2008prl,liu2011prl}. The out-of-plane current component in the Pt layer was excluded from the simulations since it does not contribute to spin current transfer into the Co layer. The charge current through the NCs also generates an Oersted (Oe) field in the Co layer. Figures \ref{fig:1}(d) and \ref{fig:1}(e) present the X- and Y-components of the Oe field resulting from a 1 mA charge current. We used the current density extracted from COMSOL for the Pt layer and the corresponding Oe field in the Co layer to conduct micromagnetic simulations with MuMax3~\cite{vansteenkiste2014aipadv}, a GPU-accelerated finite-difference solver. This solver calculates the time-dependent normalized magnetization by solving the Landau-Lifshitz-Gilbert-Slonczewski (LLGS) equation for each simulation cell. The torque exerted on the ferromagnetic layer is given by the term~\cite{mykola2018pra}
\begin{equation}
    \Vec{\tau} = \frac{\Theta_{SH} \hbar \gamma}{2M_set} \left[\Vec{m} \times \left(\Vec{e_z} \times \Vec{J_c} \right) \times \Vec{m} \right]
\end{equation}
where $\hbar$, $\gamma$, and $e$ are reduced Planck's constant, gyromagnetic ratio, and charge of an electron, respectively. $M_{s}$ and $t$ denote the saturation magnetization and thickness of the ferromagnetic layer. The injected moments in the ferromagnetic layer are oriented in the direction of $(\Vec{e_z} \times \Vec{J_c})$, and these moments are calculated for each cell, as the direction of $\Vec{J_c}$ varies from cell to cell. The FM layer is discretized into a 1024 x 1024 x 1 grid, with each cell having dimensions of 1 x 1 x 0.8 nm$^3$. The simulations utilize material parameters derived from values reported in previous studies~\cite{metaxas2007prl,durrenfel2017prb,khadka2018jap}: $M_{s} = 1.31 \times 10^6$ A/m, exchange constant $A = 22$ pJ/m, Gilbert damping $\alpha = 0.3 $, perpendicular magnetic anisotropy constant $K_\perp = 1.28 \times 10^6$ J/m$^3$, and DMI constant $D = 2$ mJ/m$^2$.

\begin{figure*}[t]
\centering
\includegraphics[width=16 cm]{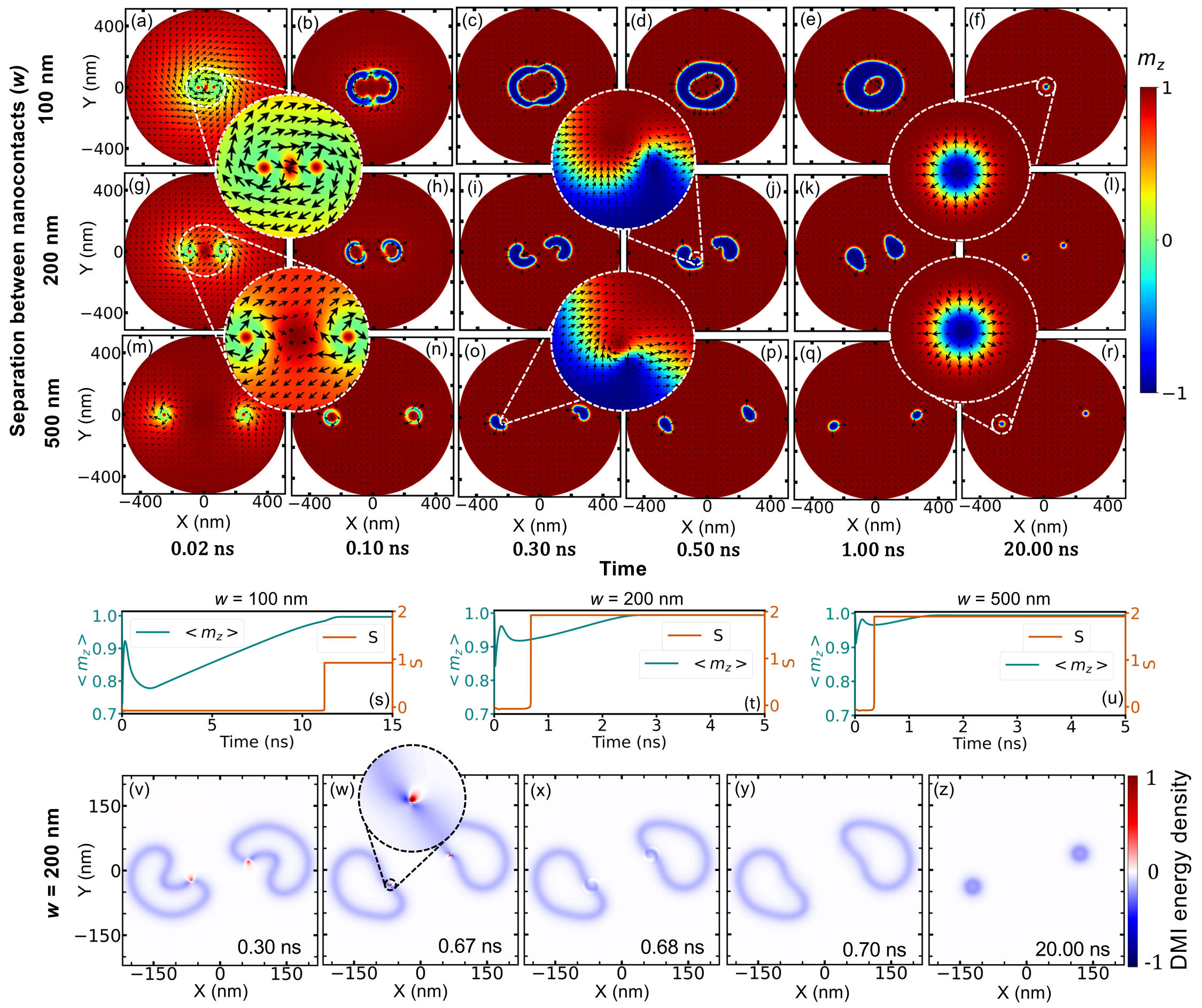}
\caption{Time snapshots illustrating the nucleation process, normalized magnetization ($m_z$), topological charge ($S$), averaged magnetization $z$-component ($<m_z>$), and DMI energy density distribution. (a-f) Time snapshots of the nucleation process for a 100 nm separation with a threshold current pulse of -66 mA, showing the formation of a ring-like domain (skyrmionium) that relaxes into a merged skyrmion state. (g-l) Time snapshots for a 200 nm separation with a threshold current pulse of -48 mA, where two independent broken ring-like states transition into a skyrmion-antiskyrmion metastable state, which stabilizes as two distinct skyrmions. (m-r) Snapshots for a 500 nm NC separation with a -40 mA threshold current pulse, showing a similar skyrmion-antiskyrmion metastable state that stabilizes into two separate skyrmions. Arrows denote the in-plane magnetization direction. (s-u) Time evolution of $<m_z>$ and $S$ for three distinct nanocontact separations: 100 nm, 200 nm, and 500 nm. (v-z) Normalized spatial distribution of DMI energy density for a 200 nm NC separation at different time steps.
}
\label{fig:2}
\end{figure*}

We begin by discussing the influence of NC separation on the nucleation and interaction mechanisms of skyrmions excited by a 20-ps current pulse, chosen to replicate the output of an experimentally realized pulse generator. Figure \ref{fig:2} (a-r) illustrates the time evolution of skyrmion nucleation for three different NC separations: $w$ = 100 nm, 200 nm, and 500 nm, each excited by a 20-ps threshold current pulse. For closely spaced nanocontacts ($w$ = 100 nm), excited by a -66 mA current pulse, we observe a strong interaction between the tilted magnetic domain configurations at the two NCs ($t$ = 0.02 ns). This interaction leads to the formation of an intermediate ring-like domain state [Fig. \ref{fig:2} (d-e)], referred to as a skyrmionium~\cite{zhang2016prb,durrenfel2017prb}, which eventually relaxes into a stabilized single-skyrmion state by $t$ = 12 ns (see Fig. \ref{fig:2} (a-f)). Initially, vortex-like structures form around each NC, while an anti-vortex-like structure appears at the center due to their strong interaction. This interaction results in the formation of two separate domains, which merge into a skyrmionium that later stabilizes into a single skyrmion, as identified by the shift in topological charge from 0 to 1 in Fig. \ref{fig:2}(s)~\cite{zhou2015ntc}. Topological charge (skyrmion number, $S$) is expressed as~\cite{zhou2015ntc,ritzmann2018ne} 
\begin{equation}
S = \frac{1}{4\pi} \int n \, dx \, dy
\end{equation}
where \( n \) is the topological charge density defined as:
\begin{equation*}
n = -\mathbf{m} \cdot \left( \frac{\partial \mathbf{m}}{\partial x} \times \frac{\partial \mathbf{m}}{\partial y} \right)
\end{equation*}

The formation of a stable, merged skyrmion state is driven by two key factors: (i) the natural tendency of the system to minimize total energy, particularly in closely spaced NCs, by reducing domain-wall energy, dipolar interactions, and the DMI energy, and (ii) the conservation of topological charge. 

When the NC separation increases to $w$ = 200 nm, the threshold current pulse magnitude decreases to -48 mA. As shown in Fig. \ref{fig:2} (g-l), the weaker interaction still excites vortex-like and antivortex-like structures ($t$ = 0.02 ns), followed by the formation of two distinct broken ring-like domain configurations ($t$ = 0.1 ns), resulting in a metastable skyrmion-antiskyrmion state~\cite{moutafis2009prb,sampaio2013ntn,buttner2017ntn,stier2017prl,ritzmann2018ne,wang2019prb,han2020jap} [Fig. \ref{fig:2} (j)]. Skyrmions and antiskyrmions exhibit distinct chiralities, with skyrmions featuring a uniform winding of spins (in or out of the plane), whereas antiskyrmions display a reversed, mirror-symmetric spin texture. The zoomed-in circle in Fig. \ref{fig:2}(j) highlights this topologically distinct, defect-like magnetization texture, which in our case arises from a complex interplay of topological charge conservation, magnetization reconfiguration, and dynamic interactions between vortex and anti-vortex states during nucleation process. Eventually, the antiskyrmion phase annihilates (see Fig. S1 in the supplementary material), nucleating two stabilized skyrmions at $t$ = 0.68 ns. Further increasing the separation to $w$ = 500 nm reduces the threshold current pulse amplitude to -40 mA. As seen in Fig. \ref{fig:2}(m-r), the nucleation process still produces similar metastable states, including skyrmion-antiskyrmion pairs ($t$ = 0.3 ns), as observed in the 200 nm case. This suggests that reduced interaction does not significantly affect the metastable states. However, at higher current pulse amplitudes, the nucleation dynamics shifts: skyrmion-antiskyrmion states disappear, and skyrmionium emerges as a new metastable state (see Fig. S2 in the Supplementary Material). In Fig. \ref{fig:2}(s-u), the sharp increase in $S$ from approximately 0 to 1 or 2 marks the nucleation of one or two stabilized skyrmions. The average $<m_z>$ component reflects the degree of magnetization reversal after applying the current pulse, as the initial state is characterized by up magnetization ($<m_z> = 1$). A constant $<m_z>$ over time signifies the presence of stable skyrmions, indicating that the $z$-component of magnetization is no longer changing.

To gain further insights into the skyrmion-antiskyrmion metastable state, we present the time snapshots of the DMI energy density analytically calculated from our simulations data for NC separation of 200 nm, following the injection of a -48 mA current pulse. The DMI energy in the continuous magnetization model is given by~\cite{bogdanov2001prl,thiaville2012europhylett}
\begin{equation}
    E_{\text{DMI}} = D \left( m_z \frac{\partial m_x}{\partial x} - m_x \frac{\partial m_z}{\partial x} + m_z \frac{\partial m_y}{\partial y} - m_y \frac{\partial m_z}{\partial y} \right)
    \label{eq:2}
\end{equation}

The energy density associated with DMI essentially governs how the magnetic moments between neighboring NCs twist relative to each other, favoring chiral structures over conventional domain walls. This DMI-driven twisting of spins can lead to a temporary alignment of opposing topological charges, resulting in a transient skyrmion-antiskyrmion state as the system seeks to locally conserve energy and topology. As shown in Fig. \ref{fig:2}(v-z), the blue and red regions in the energy distribution correspond to different chiralities, decreasing or increasing the DMI energy. The antiskyrmion state~\cite{koshibae2016ntc,sen2022pra}, characterized as an unstable elementary excitation in systems with isotropic DMI, decays over time due to Gilbert damping~\cite{stier2017prl,wang2019prb}. The DMI energy density with positive values, corresponding to the energetically unfavored region, gradually shrinks from 0.3 ns (Fig. \ref{fig:2}(v)) to 0.67 ns (Fig. \ref{fig:2}(w)) before annihilating, as seen in Fig. \ref{fig:2}(x). The formation of skyrmion-antiskyrmion magnetic texture reflects the intricate interplay between the system's topological properties and energy minimization processes.

\begin{figure}[t]
\centering
\includegraphics[width=8.5 cm]{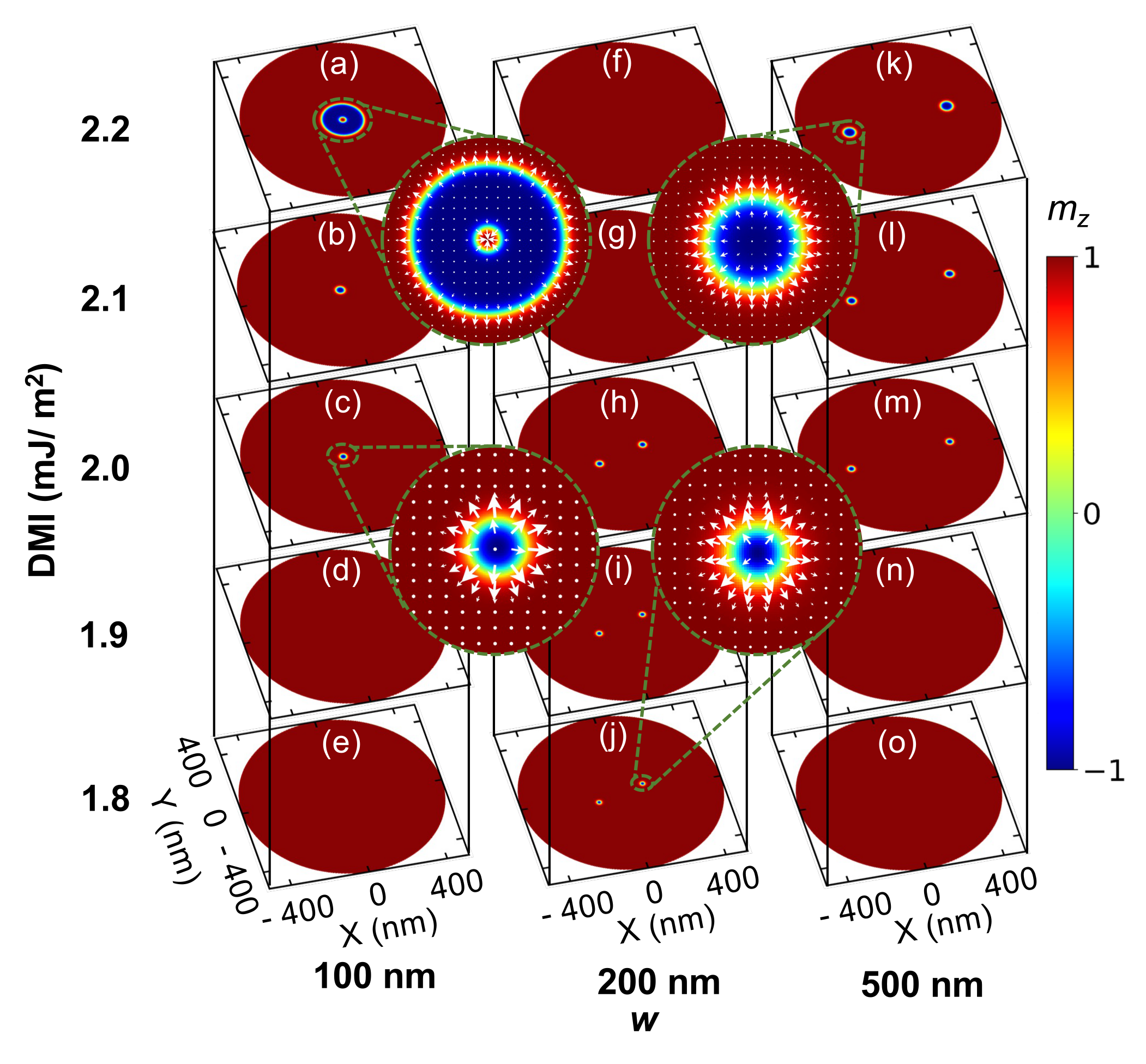}
\caption{Impact of DMI on the nucleation dynamics of skyrmions at three different NC separations. (a-e) Evolution of stable magnetization configurations with decreasing DMI for 100 nm-separated NCs under a -66 mA current pulse, showing (a) a stable skyrmionium at a DMI of 2.2 mJ/m$^2$, (b, c) stable skyrmions between 2 and 2.1 mJ/m$^2$, and (d, e) no skyrmion nucleation below 2 mJ/m$^2$. (f-j) Evolution of magnetization configurations for 200 nm-separated NCs under a -48 mA current pulse, revealing (f, g) no skyrmion nucleation for DMI $\geq$ 2 mJ/m$^2$, and (h-j) stable skyrmions between 1.8 and 2 mJ/m$^2$. (k-o) Evolution of magnetization configurations for 500 nm-separated NCs under a -40 mA current pulse, showing (k-m) stable skyrmions between 2 and 2.2 mJ/m$^2$, and (n, o) no skyrmion nucleation for DMI $\leq$ 2 mJ/m$^2$.
}
\label{fig:3}
\end{figure}

Next, we investigate the influence of DMI on the skyrmion nucleation process beneath two NCs with varying separations. As illustrated in Fig. \ref{fig:3}(a-e), for NCs separated by 100 nm and subjected to a -66 mA current pulse, a stabilized skyrmion is observed for DMI values between 2.0 and 2.1 mJ/m$^2$. When the DMI reaches or exceeds 2.0 mJ/m$^2$, an intermediate skyrmionium state forms, remaining stable~\cite{bo2020jopdap,yang2023ntc} at 2.2 mJ/m$^2$ (Fig. \ref{fig:3}(a)). Notably, skyrmion stabilization does not occur at DMI values below 2.0 mJ/m$^2$. The DMI value significantly influences the size and chirality of the skyrmions; higher DMI values result in larger skyrmions~\cite{sampaio2013ntn} or skyrmionium~\cite{yang2023ntc} due to stronger twisting of spins and a reduced energy barrier. Conversely, lower DMI values lead to smaller and less stable skyrmions. Furthermore, as DMI increases, the threshold current required for skyrmion nucleation decreases, and the energy landscape becomes more favorable for skyrmionium formation rather than isolated skyrmions. For NCs separated by 200 nm and subjected to a -48 mA current pulse (Fig. \ref{fig:3}(f-j)), no skyrmion nucleation is observed at DMI values above 2.0 mJ/m$^2$. In this case, stable skyrmions form for DMI values between 1.8 and 2.0 mJ/m$^2$. Similarly, for NCs separated by 500 nm and exposed to a -40 mA current pulse, stable skyrmions appear for DMI values between 2.0 and 2.2 mJ/m$^2$, while no nucleation occurs for DMI values below 2.0 mJ/m$^2$. The contrasting behaviors in the 200 nm and 500 nm NC separations arise from the interplay between DMI strength and the applied current pulse amplitude. These findings indicate the critical role of both DMI and the applied threshold current in modulating the interactions between vortex-like states during the nucleation process.

\begin{figure}[ht]
\centering
\includegraphics[width=8 cm]{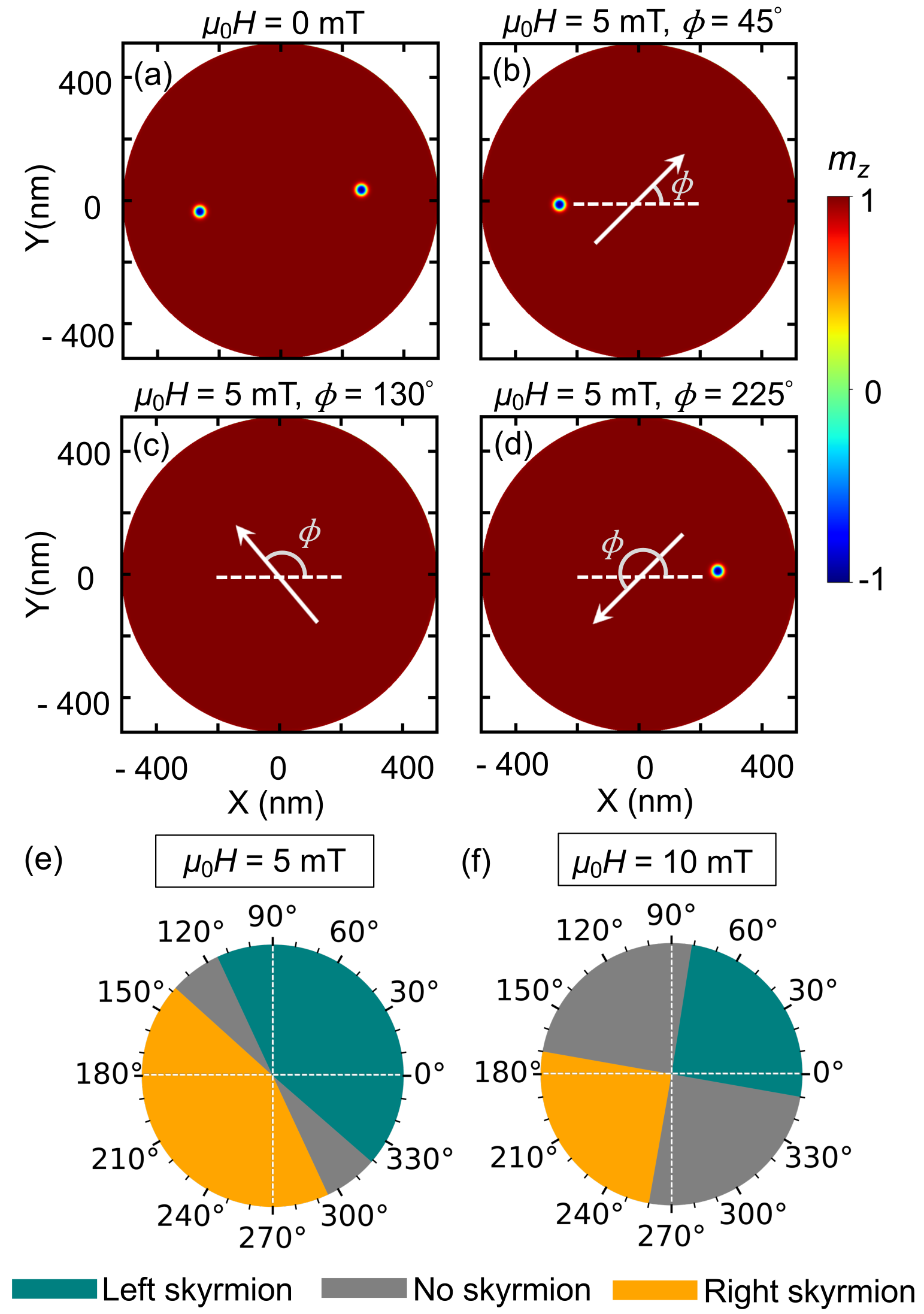}
\caption{Effect of applied in-plane magnetic field on skyrmion stability for potential applications. (a) Two skyrmions are stabilized in the absence of an applied magnetic field. (b) The right skyrmion fails to nucleate under a 5 mT in-plane field oriented at an angle of $\phi$ = 45$^{\circ}$, highlighting selective skyrmion control. (c) No skyrmion nucleation occurs under a 5 mT in-plane field oriented at $\phi$ = 130$^{\circ}$, indicating field-dependent suppression. (d) The left skyrmion fails to nucleate under a 5 mT in-plane field oriented at $\phi$ = 225$^{\circ}$. (e, f) Angular color plots illustrate the magnetic field orientations that selectively control nucleation of the left or right skyrmion, or completely suppress nucleation, under 5 mT and 10 mT fields. 
}
\label{fig:4}
\end{figure}

Finally, we investigate the influence of a weak applied in-plane magnetic field on skyrmion stability underneath two NCs. We observe that the orientation ($\phi$) of the in-plane magnetic field enables the control over the nucleation and suppression of either one or both skyrmions. As illustrated in Fig. \ref{fig:4}(a-d), the effects of applied in-plane magnetic fields on NCs separated by a distance of 500 nm are demonstrated for three different orientations. The fields are continuously applied, while a -40 mA pulse lasting 20 ps induces the transition from the initial up-magnetized state to the final state, depending on the in-plane orientation of the applied fields. In Fig. \ref{fig:4}(a), we observe a stable skyrmion pair in the absence of an applied field. When a 5 mT field is applied at an angle of 45$^{\circ}$ (Fig. \ref{fig:4}(b)), only the left skyrmion stabilizes. The suppression mechanism of the right skyrmion under an in-plane field at an angle of $\phi$ = 45$^{\circ}$ is illustrated in the supplementary material (Fig. S3) for two different applied fields, 5 mT and 10 mT. We demonstrate that the presence of a weak in-plane field does not annihilate the antiskyrmion component of the skyrmion-antiskyrmion texture, resulting in the suppression of the skyrmion state. Furthermore, Figs. \ref{fig:4}(c-d) depict stable magnetization configurations, showing no skyrmion at 130$^{\circ}$ and only the right skyrmion at 225$^{\circ}$ for an applied field of 5 mT. Upon increasing the in-plane field strength to 10 mT, we observed similar behavior across different orientations, with an expanded angular region of skyrmion suppression. The relationship between the controlled nucleation of skyrmions and the varying $\phi$ for in-plane magnetic fields of 5 mT and 10 mT is summarized in Figs. \ref{fig:4}(e) and (f), respectively. A comparative analysis reveals that the region with no skyrmion nucleation (gray) expands as the field strength increases. In addition, symmetrical behavior is observed between left skyrmion nucleation (green) and right skyrmion nucleation (orange) as a function of $\phi$. We reproduced the similar in-plane field-dependent control of skyrmion nucleation for a 200 nm NC separation (see Fig. S4 in the supplementary material). This field-dependent behavior can be leveraged in applications such as in-plane field-controlled bits, field-programmable logic~\cite{yan2021pra}, and skyrmion switches~\cite{schott2017nanolett}.

In conclusion, we systematically investigated the nucleation and interaction dynamics of magnetic skyrmion pairs underneath two 30 nm NCs, excited by a 20-ps current pulse via SOT in an extended bilayer thin films. By adjusting the NC separation, DMI strength, and current pulse amplitude, we demonstrated effective control over the interaction of magnetization configurations, enabling the tunable nucleation of distinct metastable states. For closely spaced NCs and higher DMI, skyrmionium states emerge, while at greater separations, a transient skyrmion-antiskyrmion pair is observed. Furthermore, the skyrmion-antiskyrmion metastable state shows high sensitivity to the orientations of weak in-plane magnetic fields, facilitating field-dependent manipulation of stable skyrmion states. These findings highlight the potential of skyrmion-antiskyrmion states for applications in skyrmion-based spintronic devices.

\section*{Supplementary Material} 
See the supplementary material for additional details.
\section*{Acknowledgments}
This work is supported by the Faculty Initiation Grant (No. IITR/SRIC/3648/FIG) sponsored by SRIC, IIT Roorkee, and the SRG/2023/002487 project funded by the Science \& Engineering Research Board (SERB), Department of Science \& Technology, Government of India. HP acknowledges the financial support from the Ministry of Education, Government of India. We thankfully acknowledge the Institute Computer Center (ICC), IIT Roorkee, for providing a high-end computational facility to run simulations.

\section*{Author declarations}
\textbf{Conflict of Interest}

The authors have no conflicts to disclose.

\section*{Data availability}
The data that support the findings of this study are available upon reasonable request from the authors. 


%

\end{document}